\documentclass[preprint,3pt,doublecolumn]{elsarticle}

\usepackage{amssymb, amsthm}

 \biboptions{comma,square}

\journal{Journal of Computational Physics}

\begin{document}

\begin{frontmatter}



\title{Molecular dynamics beyonds the limits: massive scaling on 72 racks of a BlueGene/P and supercooled glass transition of a 1 billion particles system}


\author{Nicholas Allsopp$^1$}
\address{$^1$Kaust Supercomputing Laboratory, King Abdullah University of Science and Technology (KAUST), Thuwal 23955-6900, Saudi Arabia}

\author{Giancarlo Ruocco$^2$}
\address{$^2$Dept. of Physics, Sapienza University of Rome, P.le A. Moro 2, 00184 Rome, Italy}

\author{Andrea Fratalocchi$^{3}$\corref{cor1}}
\address{
$^3$PRIMALIGHT, Faculty of Electrical Engineering; Applied Mathematics and Computational Science, King Abdullah University of Science and Technology (KAUST), Thuwal 23955-6900, Saudi Arabia
}
\ead{andrea.fratalocchi@kaust.edu.sa}
\ead[url]{www.primalight.org}

\begin{abstract}
We report scaling results on the world's largest supercomputer of
our recently developed Billions-Body Molecular Dynamics
($\mathsf{BBMD}$) package, which was especially designed for
massively parallel simulations of the atomic dynamics in
structural glasses and amorphous materials. The code was able to
scale up to 72 racks of an IBM BlueGene/P, with a measured 89\%
efficiency for a system with 100 billion particles. The code
speed, with less than 0.14 seconds per iteration in the case of 1
billion particles, paves the way to the study of billion-body
structural glasses with a resolution increase of two orders of
magnitude with respect to the largest simulation ever reported. We
demonstrate the effectiveness of our code by studying the
liquid-glass transition of an exceptionally large system made by a
binary mixture of 1 billion particles.

\end{abstract}


\end{frontmatter}


\section{Introduction}
\label{preface}

Recently, contributing to the 2009 annual survey of the EDGE
foundation \cite{edge} entitled "What will change everything" \cite{sejnowski94:_comput_brain}, T.
Sejnowski foresees that the computers will be the "microscopes of
the future" and, specifically, that are computers that "have made
it possible to localize single molecules with nanometer precision
and image the extraordinary complex molecular organization inside
cells". In his contribution Sejnowski was recognizing the
importance of automatic and computerized control of laser beam and
image analysis in modern optical microscopy. However, more importantly, computers
can be also considered as "future
microscopes" for the capability of simulating at the atomic scale
the behavior of matter and biological systems. To reach
this goal, which was even unthinkable up to some decades
ago, it is necessary to develop software able
to simulate the newtonian dynamics of a large number of atoms
(order $10^{12}$, the "number" of atoms in a cell) for long enough
time (order microsecond, or $10^{10}$ time steps) \cite{rapaport04:_art_of_molec_dynam_simul,marx09:_ab_initio_molec_dynam,frenkel01:_under_molec_simul_secon_edition}. The state-of-the art supercomputers, on the hardware side, and the parallel
simulation codes, on the software one, are nowadays on the way to
get this result. However, before facing the ultimate problem of simulating the complexity of
life (micro-)organisms, one should validate and optimize codes that
simulate hard (physical and chemical) systems. Even if smaller than a micro-organism,
these systems encompass problems which are extremely demanding in terms of computational resources, with e.g., simulation boxes containing pico-molar quantity of matter and with characteristic times of $0.1$ microseconds. Molecular Dynamics (MD) \cite{rapaport04:_art_of_molec_dynam_simul} is expected to be the key to efficiently solve these type of problems.
As each generation of computers is introduced, in fact, larger and longer
simulations are allowed to be run, thus producing better results
which answer a variety of new scientific questions. The latest
generation of terascale and petascale supercomputers (see e.g.,
\cite{httpnscc,httpornl,httpjuge}), in particular, holds the promise to enable
the development of more realistic and complex interactions, as
well as the study of systems made by a very large number ($\approx 10^{10}$) of
particles. In the most common paradigm used today, called
massively parallel processing, the typical size of computer
clusters ranges from several thousand to hundreds of thousands of
processing core units. In these architectures, a high degree of
parallelization is essential to efficiently utilize the
computational power available. While in principle a high level
parallelization strategy works fine for small to medium size
supercomputer clusters, tuning for specific architectures can be
the key to achieve huge scaling performance. The design concepts
of today's processors, in fact, are markedly different from one
system to another, and it is necessary to prepare codes having
specific architectures in mind in order to optimize both the speed
and the bandwidth of memory access, which is typically slow if
compared to the processor's frequency.\\
With reference to molecular dynamics, several methods have been
published in the past which incorporate different degrees of
parallelism \cite{rapaport04:_art_of_molec_dynam_simul}. To date,
MD scaling has been demonstrated up to ten thousand of cores, with
a speed of about 7 iterations per second for an ensemble of 1
billion particles running on 65536 cores of a BlueGene/L
\cite{kadau06:_molec_dynam_comes_of_age}. For several applications
of molecular dynamics, such as the study of structural glasses
\cite{elliot83:_physic_of_amorp_mater} where a typical simulation
requires $10^{6-7}$  time steps, this speedup is still too small
to perform practical calculations. In the study of glasses, and in
the more general area of amorphous materials, molecular dynamics
simulations are extremely important as they are able to glimpse
the system dynamics at spatial scales between $1-100$ nm, a range
that is completely inaccessible with experimental apparatus
\cite{sette98:_dynam_of_glass_and_glass,shintani08:_univer_link_between_boson_peak}.
However, due to such speed bottleneck problems, present MD studies
on glass have been limited to a number of particles (10 million
\cite{monacoa09:_anomal_proper_of_acous_excit}) unable to have
simulation boxes large enough to capture the interesting
phenomenology at the micron-scale. Therefore, if on one hand it is
necessary to improve the scaling of general MD codes at least two
orders of magnitude, in order to effectively use the computing
power available, on the other it is also important to focus on
specific research fields, such as the context of amorphous
materials, which are now asking for new levels of performances.\\
In this article we report the results of our recently developed
Billions Body Molecular Dynamics ($\mathsf{BBMD}$) package, and
demonstrate its effectiveness in the study (as case study) of
structural glasses by analyzing the glass formation of an
exceptionally-large particles system. The $\mathsf{BBMD}$ code was
able to scale on the whole  $294912$ cores of the BlueGene/P
system at the J\"ulich Supercomputing Centre, which constitutes
the world's largest supercomputer available characterized by 72
racks of an IBM BlueGene/P. In such an extreme scaling test, the
$\mathsf{BBMD}$ code showed an efficiency of about 90\%, and a
overall speed of 2 seconds x iteration with 100 billion particles.
These results pave the way to the study of very large systems. In
order to demonstrate the applicability of our code to the field of
amorphous materials, we performed a controlled temperature MD
simulation of a system made by 1 billion particles, and studied
the supercooled dynamics of the liquid state by varying the
temperature in the range $T\in[2,10^{-4}]$. This simulation has
been performed on the Shaheen supercomputer, hosted at KAUST
University and constituted by 16 racks of an IBM BlueGene/P. This
paper is organized as follows. We begin our analysis by discussing
the structure of the $\mathsf{BBMD}$ code (Sec. \ref{code}),
reviewing both parallelizations and optimization strategies. In
Sec. \ref{scaling}, we describe $\mathsf{BBMD}$ scaling results
obtained at the J\"ulich supercomputing center. We report the code
speedup for molecular systems of different size, ranging from 1
billion to 100 billion particles. Communication workload versus
calculation execution time is also studied. In Sec. \ref{glassy},
finally, we investigate the glassy dynamics of a 1 billion
particles system made by a binary mixture of soft-spheres. 
\begin{figure}
 \includegraphics[width=12 cm]{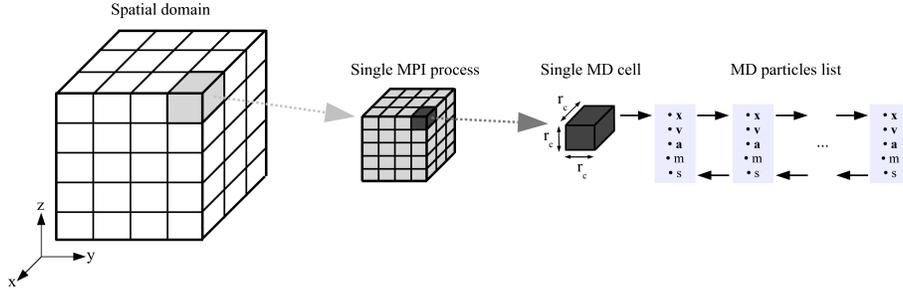}
 \caption{\label{f1} Main structure of the $\mathsf{BBMD}$
  parallel code: (right) the simulation domain is spatially
  decomposed into rectangular boxes, each defining a single
  MPI process (center) that, in turn, is constituted by
  several cubic MD cells of side $r_c$ (left). Each MD cell
  contains a bidirectional list with all particle informations:
  position $\mathbf{x}$, velocity $\mathbf{v}$, acceleration
  $\mathbf{a}$, mass $m$ and species $s$. }
 \end{figure}

\section{The $\mathsf{BBMD}$ code}
\label{code} The $\mathsf{BBMD}$ code is a highly optimized,
parallel C++ MD code for Lennard-Jones particles systems, designed
to scale on machines characterized by hundreds of thousands of
processors, such as the latest generation of IBM BlueGene/P
supercomputers. Much effort was taken to balance design simplicity
and code speed, while optimizing at the same time both memory
requirements and cache efficiency. In the following, we briefly
describe the main structure of the code.
\subsection{Interaction potential}
$\mathsf{BBMD}$ is originally designed to support two main classes
of short range interaction potentials:\\
\begin{itemize}
\item The Lennard-Jones potential
\begin{equation}
\label{lj0}
V(r_{ij})=4\epsilon_{\alpha\beta}\bigg[\bigg(
\frac{\sigma_{\alpha\beta}}{r_{ij}} \bigg)^{12}-\bigg(
\frac{\sigma_{\alpha\beta}}{r_{ij}} \bigg)^{6}  \bigg]
\end{equation}
\item The soft-sphere potential
\begin{equation}
\label{sc0}
V(r_{ij})=4\epsilon_{\alpha\beta}\bigg( \frac{\sigma_{\alpha\beta}}{r_{ij}} \bigg)^{12}
\end{equation}
\end{itemize}
being $\sigma$ and $\epsilon$ two tensors that parametrize the
potentials, and $r_{ij}$ the distance between particles $i,j$ of
species $\alpha$ and $\beta$, respectively. Since Eqs.
(\ref{lj0})-(\ref{sc0}) converge rapidly to zero beyond
$r_{ij}\approx ||\sigma||$, it is wasteful to consider an
interaction between two particles at a long distance. A standard
choice in MD is therefore to truncate the potentials beyond the
distance $r_c$, in order to increase calculation's speed. To avoid
any problem due to the discontinuity of $V(r)$ at $r=r_c$, we
replace Eqs. (\ref{lj0})-(\ref{sc0}) with the following potential:
\begin{equation}
\label{mod0}
V^*(r)=\bigg[V(r)-V(r_c)-(r-r_c)\frac{dV(r)}{r}|_{r=r_c}\bigg][1-\Theta(r_c)]
\end{equation}
with $\Theta(x)$ being the Heaviside function. The modification
(\ref{mod0}) applies across the entire interaction range, and the
overall spatial domain is decomposed into MD cubic cells of $r_c
\times r_c \times r_c$ volume (Fig. \ref{f1}).

\subsection{Grid-search method}
To perform the time evolution of the system, forces between
particles have to be calculated and particle pairs whose distance
is below the cutoff range $r_c$ have to be found. For this task,
we adopt an $O(N)$ linked-list method, with an inexclusive
\cite{rapaport04:_art_of_molec_dynam_simul} grid that allows more
than one particle to occupy a single cell. Newton's third law is
then applied in order to halve the number of neighbors to be
checked in the calculation of the forces. In order to guarantee
optimal memory efficiency, we developed a bidirectional list
structure that contains all the information of the particles:
position $\mathbf{x}$, velocity $\mathbf{v}$, acceleration
$\mathbf{a}$, mass $\mathbf{m}$ and species index $\mathbf{s}$
(Fig. \ref{f1}). The double pointer mechanism allowed an efficient
implementation of memory-related operations (e.g., movement of a
particles on a different cell or on a different processor) without
deleting or creating memory locations but just by moving pointers,
which is very fast. Another advantage of such memory structure is
that it automatically links together particles which are close in
space, thus improving speed efficiency in all search operations.
To maintain memory requirement as low as possible, we do not
employ any bookkeeping method, such as the neighboring particle
list \cite{rapaport04:_art_of_molec_dynam_simul}, which are too
memory demanding for billion sized particle systems.

\subsection{Parallelization}
\subsubsection{Parallelization scheme}
$\mathsf{BBMD}$ has been parallelized with the domain
decomposition strategy, also known as spatial decomposition, where
the simulation box is divided into subdomains and each domain is
assigned to each processor (Fig. \ref{f1}). The Message Passing
Interface (MPI) standard is then employed to handle all parallel
communications among processes. This type of parallelization has
the advantage to require only nearest-neighbor communications,
with a few limited global collective operations, and it is
therefore well suited for very large supercomputer clusters such
as the IBM BlueGene series.
\subsubsection{Parallel force calculation}
\label{pfc} $\mathsf{BBMD}$ employs the velocity-verlet time
marching algorithm \cite{rapaport04:_art_of_molec_dynam_simul},
which is a standard in MD algorithms due to its robustness and
accuracy in maintaining conserved quantities, such as the energy
and the momentum. In this evolution scheme, the hot spot of the
algorithm is the computation of the forces exerted among
particles. To perform this calculation, two parallel operations
are required: (i) moving particles that stray from the originally
associated process and (ii) exchanging particles that belong to
borders crossing different processors. In $\mathsf{BBMD}$,
particular care has been taken in the design of (i) and (ii) in
order to overlap communications and calculations to the maximum
extent possible, while optimizing speed and cache efficiency. More
specifically, the parallel communication starts with the operation
(i) and then proceeds with (ii). In both steps, a one dimensional
array containing particles properties (i.e., $\mathbf{x}$,
$\mathbf{v}$, $\mathbf{a}$, $m$, $s$) needs to be constructed, and
the number of particles to be sent to neighboring processors has
to be calculated. In $\mathsf{BBMD}$, these two operations are
overlapped with $\mathsf{send/recv}$ MPI communication routines in
order to minimize communication time with respect to calculation.
The task (i) begins by tagging all the particles that belongs to
different processors, and sorting them on-the-fly thus minimizing
access times in subsequent MPI communications. This is done by
exploiting the bidirectional nature of the particle list. In
particular, each particle that needs to be sent to a different
process is first moved on the tail of its cell list with an $O(1)$
operation, and then tagged by inverting its mass sign. Such
tagging procedure avoids the use of an external index and
increases the memory efficiency of the code. Besides that, this
method automatically groups tagged particles together, and allows
to access them sequentially with $O(1)$ operations when MPI
$\mathsf{send/recv}$ communications are performed. Once the task
(i) has been completed, the exchanging of the nearest-neighbor is
done with the standard ghost-cell approach
\cite{marx09:_ab_initio_molec_dynam}. In both (i) and (ii) tasks,
one single MPI process is required to communicate with his 26
neighbors. Although being characterized by nearest neighbors
communication only, a naive implementation of this algorithm
requires 26 different communications and results in poor scaling
performances when a large number of processor is employed.
However, by taking advantage of the domain decomposition strategy
employed in $\mathsf{BBMD}$,  it is possible to reduce the number
of total communications to just 6. This is achieved by properly
enlarging the communication window, and in particular by
exchanging part of the ghost cells during each MPI communication
(see e.g., \cite{frenkel01:_under_molec_simul_secon_edition} for
more details).
\begin{figure}
\centering
 \includegraphics[width=9 cm]{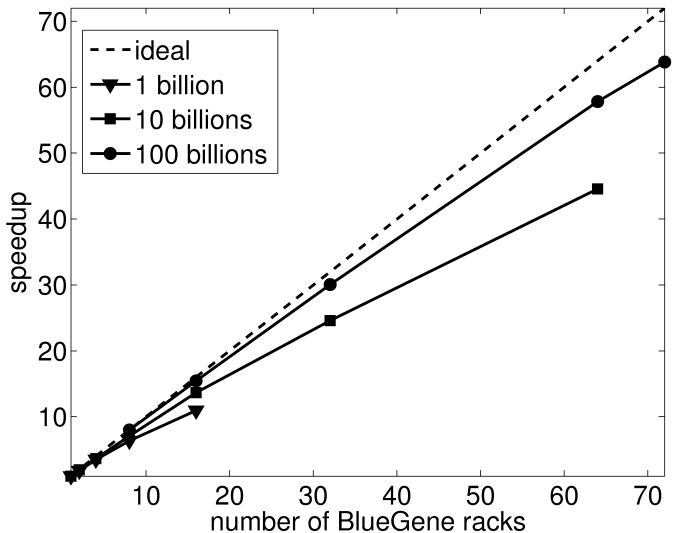}
 \caption{\label{f2} $\mathsf{BBMD}$ strong scaling results: code speed versus number of processors.}
 \end{figure}
\subsubsection{BlueGene specific optimizations}
In the calculation of the $\mathsf{sqrt}$ function, required by the computation of Eq. (\ref{mod0}), we employ the square root reciprocal BlueGene function $\mathsf{frsqrte}$, coupled with two Newton-Rapson iterations. More specifically, we replaced the code segment:
\begin{verbatim}
rij=sqrt(r2);
\end{verbatim}
with:
\begin{verbatim}
rij = frsqrte(r2);
rij = ((0.5 * rij) * (3.0 - r2 * (rij * rij)));
rij = ((0.5 * rij) * (3.0 - r2 * (rij * rij)));
rij = rij * r2;
\end{verbatim}
This optimization results in a speed increment of about 7\%.

\section{$\mathsf{BBMD}$ scaling results}
\label{scaling} 
\begin{figure}
\centering
 \includegraphics[width=9 cm]{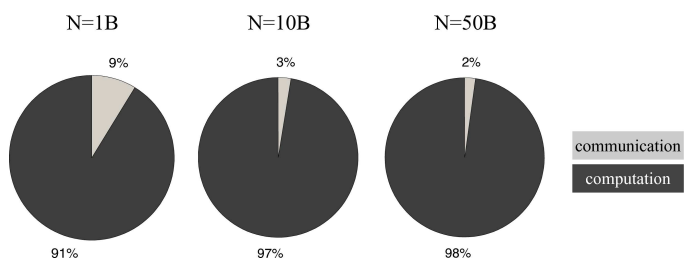}
 \caption{\label{f4} $\mathsf{BBMD}$ computation versus communication workload for system with increasing particles number N. The measurements have been performed with simulations of 1000 time steps on 16 racks.}
 \end{figure}

The evaluation of $\mathsf{BBMD}$ code
performances has been carried out on the Jugene system at the
J\"ulich Supercomputing Center \cite{httpjuge}, which is composed
of $294912$ cores (or 72 racks) of an IBM BlueGene/P with total
peak performance of 1 PFlops. The test suite consisted of a series
of canonical molecular dynamics simulations of a $20:80$ binary
mixture of soft-spheres with the following parameters (here all
the units are to be intended as normalized MD units
\cite{frenkel01:_under_molec_simul_secon_edition}):
$m_i=1$, $\sigma_{11}=1$, $\sigma_{12}=0.8$, $\sigma_{22}=0.88$, $\epsilon_{11}=1$, $\epsilon_{12}=1.5$, $\epsilon_{22}=0.5$. In the canonical evolution, the temperature $T$ has been fixed to $T=0.5$, and the density $\rho$ to $\rho=1.2$ in order to maintain the system in the liquid state with all the particles randomly fluctuating among different processors. Figure \ref{f2} displays $\mathsf{BBMD}$ strong scaling results for systems characterized by $1,10$ and 100 billion particles. Although the memory footprint of $\mathsf{BBMD}$ permits the number of particles to be increased by up to two orders of magnitude, we concentrated on a range where the code speed was sufficiently fast to implement realistic calculations. In the smallest case of a system with 1 billion of particles, $\mathsf{BBMD}$ was able to scale well beyond 10 racks and obtained and efficiency of 68\% on 16 racks when compared to a single rack. When the number of particles was increased from 1 billion to 10 billion, 70\% of efficiency was achieved between 1 rack and 64 racks. This number was improved even further for systems containing 100 billion particles. In this configuration, $\mathsf{BBMD}$ reached an efficiency of 89\% on 72 racks (or 294912 cores) when compared to 8 racks. \\
\begin{figure}
\centering
 \includegraphics[width=9 cm]{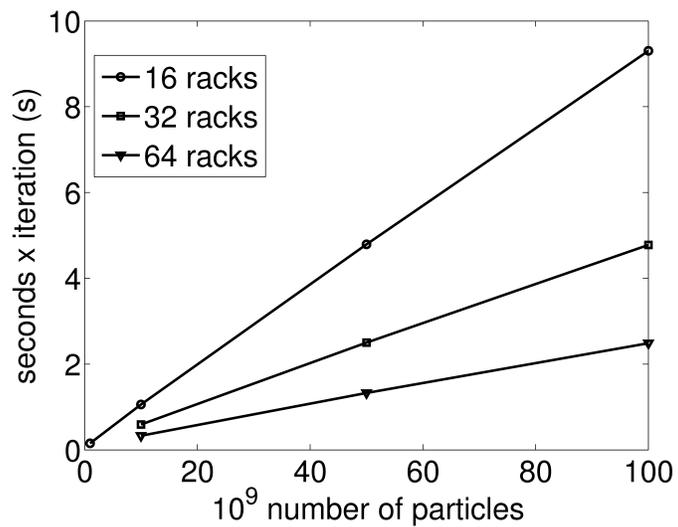}
 \caption{\label{f3} $\mathsf{BBMD}$ weak scaling results: code speed versus number of particles.}
 \end{figure}
To investigate the weak scaling performances of the code, we
perform a series of runs with a fixed number of cores and with
system sizes varying between $N=1$ billion and $N=100$ billion.
Figure \ref{f3} shows the result of such analysis. When the
particles are increased up to three orders of magnitude, the
$\mathsf{BBMD}$ code shows a perfect linear $O(N)$ complexity. The
improved scaling is due to the proportion of computation with
respect to communication time, which appreciably increases when
the number of particles grow (Fig. \ref{f4}).

As seen in Fig. \ref{f4}, the communication impact over computation
is practically negligible and the overall execution time is dominated
by calculations. This is not only the result of the three dimensional
spatial domain decomposition employed, which significantly reduces the
volume of each MPI process and the number of elements to be exchanged
during each iteration, but also relies on the specific overlapping
strategy used in the parallel calculation of the forces (Section \ref{pfc}),
which minimizes $\mathsf{MPI\_ Waitall}$ times.\\
In the case of a system with 1 billion of particles, we measured a
code speed of 0.14 seconds x iteration, which is sufficiently
small to enable the study of billion-body structural glasses. Such
a system size is two orders of magnitude higher than the largest
glass simulation ever reported, and will improve up to two orders
of magnitude the resolution of MD measurements on glassy dynamics
\cite{monacoa09:_anomal_proper_of_acous_excit}.

\section{A glass transition in a 1 billion binary mixture of soft spheres}
\label{glassy}

\subsection{A few words about the glass phase}
When the temperature of a liquid is reduced below the melting
temperature, two possible physical phenomena may occur: either the
system undergoes \emph{crystallization} -a first order phase
transition- where the final ordered configuration is the
thermodynamically stable phase, or else the liquid may become
\emph{supercooled} and get dynamically arrested into a disordered
solid represented by a glass. Despite a large scientific
production, the phenomenology of the glass state is still far from
being completely understood, and many aspects are yet unknown. As
an example, it is not yet clear whether the dynamical arrest is a
genuine thermodynamic phase transition, a kinetic phase transition
er something else, as the real world counterpart of a phase
transition taking place in the trajectories' phase space as
recently suggested by Chandler and co-workers \cite{Chandler}. In
the last decades, in particular, several different theories have
been suggested and various diverse analysis have been pursued
\cite{sette98:_dynam_of_glass_and_glass,shintani08:_univer_link_between_boson_peak,
angell95:_format_of_glass_from_liquid_and_biopol,tarjus01:_ammin_and_rheol,debenedetti01:_super_liquid_and_glass_trans,cavagna09:_super_liquid_for_pedes}.\\
Notwithstanding the wide diversity of views, a consensus exists
over the consideration that a glass transition is a dynamical
crossover through which a viscous liquid falls out from
equilibrium and becomes solid on the experimental time scale. Such
a process is manifested in a gradual change in the slope of the
volume (or other extensive thermodynamic variables such as the
entropy or the entalphy) at a specific temperature $T_g$, which
defines the \emph{glass-transition} temperature. A fundamental
property of glasses is the existence of a high degree of
\emph{frustration} in their ground state energy configuration
\cite{MPVBook}. Frustration, in turn, is manifested by the
existence a huge number of minima of equivalent energy (metastable
states). When the temperature decreases below a specific
threshold, the energy barriers of the various minima becomes
sufficiently large to trap the dynamics in the configuration space
and let it explore only a subset of the available iso-energy
surface. The result of this dynamical arrest is a disordered solid
that defines the glass phase
\cite{cavagna09:_super_liquid_for_pedes}.
\begin{figure}
\centering
 \includegraphics[width=8 cm]{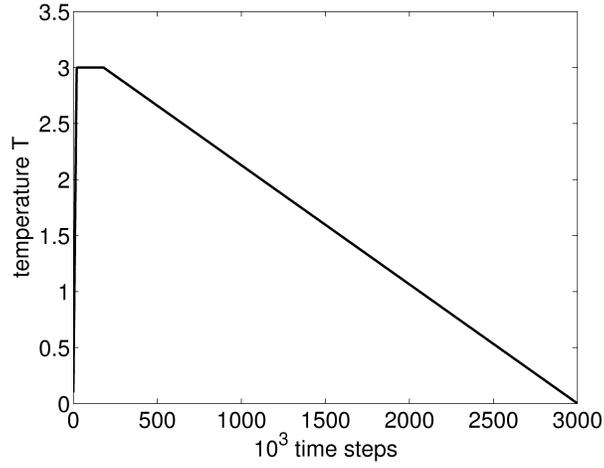}
 \caption{\label{f5} Temperature $T$ curve as a function of the time step employed in the study of the liquid-glass transition in a 1 billion particles system.}
 \end{figure}

\begin{figure}
\centering
 \includegraphics[width=9 cm]{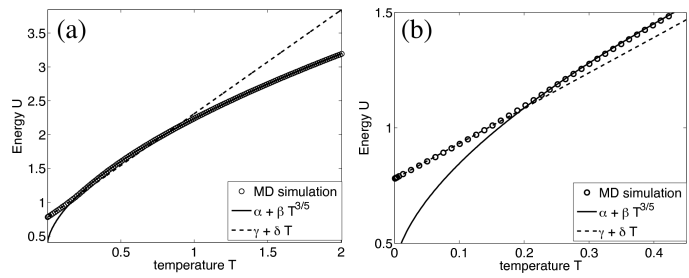}
 \caption{\label{f6} Caloric curves $U(T)$ obtained from MD simulations (circle markers) and relative least square fit (solid and dashed lines). Figure (b) is an enlarged section of (a) near the glass temperature $T_g$.}
 \end{figure}

\subsection{Sample setup and simulation results}
\begin{figure}
\centering
 \includegraphics[width=5 cm]{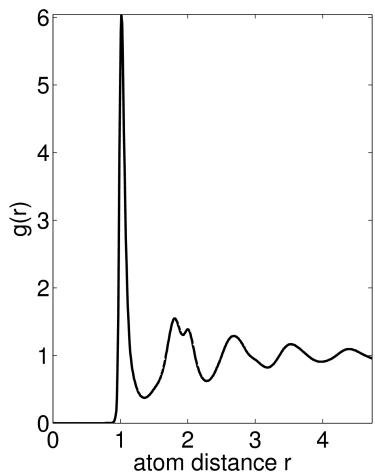}
 \caption{\label{f8} The radial distribution function $g(r)$ of the glass at $T=10^{-3}$.}
 \end{figure}

It is worth to mention that on cooling a liquid, if
crystallization is avoided, the atomic dynamics is characterized
by a relaxation time (naively, one can thing to the typical time
needed to change the inherent "configuration") that increase
following an Arrhenius law in strong glass forming systems, and
even faster in fragile glass forming systems \cite{Angell,
Ruocco}, reaching the value of 100 s at the glass transition
temperature. A supercooled liquid above, but close to, the glass
transition temperature can be considered at equilibrium only if
its atomic dynamics is investigated for log time, longer that the
relaxation time, a macroscopic time. As a consequence, during a MD
run, necessarily lasting for time much shorter than the relaxation
times close to the glass transition, one falls out of equilibrium
as soon as the temperature became that temperature where the
relaxation time correspond to the simulation times. As a
consequence, in a MD simulation the temperature where the system
becomes trapped in a specific inherent structure is largely higher
than the real glass transition temperature. In a molecular
dynamics simulation, a liquid-glass transition can be observed as
a continuous transition of the extensive thermodynamic parameter
as energy or specific volume.\\
To study this transition, we employed the same binary mixture used
for our benchmark suite in Section \ref{scaling}: the high density
$\rho=1.2$ and the $20:80$ highly asymmetric mixture
configuration, in fact, neglects the existence of a well defined
solid phase for the system and the low temperature ground state
becomes highly frustrated
\cite{angelani04:_saddl_and_softn_in_simpl_model_liquid}. We
therefore expect the observation of a glass transition in the
dynamics as soon as the temperature is decreased to a small enough
value. Figure \ref{f5} shows the temperature curve employed for
our MD simulation. At the beginning, we rapidly increase the
system temperature to a high value $T=3$. In this state, the two
particles species have sufficient kinetic energy to freely diffuse
over the whole simulation box. After $200000$ time steps, the
temperature starts to decrease by following a slowly-varying
linear curve, whose duration is of $3\cdot 10^6$ time steps. In
order to guarantee stability and energy conservation over the
whole simulation, we adopt a time $t$ resolution of $\delta
t=10^{-3}$ s. The cutoff range has been set to $r_c=1.2$, which
guarantees a sufficiently large interaction distance to observe
the formation of a supercooled liquid.
 \begin{figure}
\centering
 \includegraphics[width=7 cm]{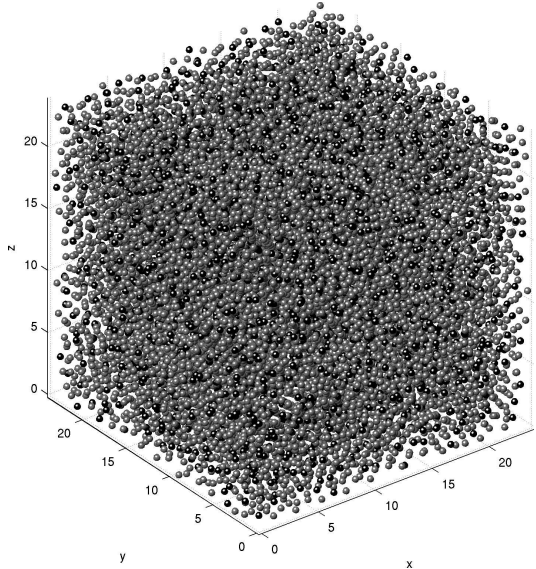}
 \caption{\label{f7} A $16000$ particles portion of the resulting glass at $T=10^{-3}$. The black particles are of species $1$, the gray belongs to species $2$.}
 \end{figure}
Figure \ref{f6} displays the caloric curve of the system, and
shows the behavior of the potential energy $U$ versus the
temperature $T$. For temperature values well above $T_g=0.2$, the
system is in the liquid state and follows a continuous  power law
curve with $T=T^\frac{3}{5}$, as found from a nonlinear least
square fit procedure applied on the energy $U$ and as expected
from the theory of Tarazona \cite{Tarazona}. As we progressively
reduce the temperature beyond the value $T_g$, we observe the
appearance of a continuous transition, characterized by a
radically change in the derivative of $U$ with respect to $T$. For
temperature values below $T_g$, which defines the glass
temperature \cite{elliot83:_physic_of_amorp_mater}, the energy $U$
varies linearly with the temperature and the system gets trapped
into an arrested phase with all the particle oscillating almost
harmonically around their equilibrium configuration. Figure
\ref{f7} displays a $16000$ particles portion of the billion-body
solid formed at $T=10^{-3}$. As expected from the thermodynamic
analysis based on the caloric curve, the solid configuration
reached is that of a structural glass, whose particles are
randomly arranged in space. The structural properties of the glass
are analyzed by calculating the radial distribution function
$g(r)$, defined as:
\begin{equation}
g(r)=\frac{2V}{N^2}\bigg\langle \sum_{i<j}\delta(r-r_{ij})\bigg\rangle,
\end{equation}
being $V$ the sample volume,  $N$ the total number of particles,
$r_{ij}$ the distance between particles $i$ and $j$ and $<...>$
denoting an ensemble average. The radial distribution function
describes the spherically averaged local organization around a
specific atom, and it measures the probability of finding an atom
at a distance $r$ from a given particle. Figure \ref{f8} displays
the $g(r)$ for the billion-body glass obtained at $T=0.001$. The
glass is characterized by a sharp peak at $r\approx 1$, which
yields the average minimum interparticle distance at equilibrium,
followed by two broader peaks and a series of oscillations of
decreasing amplitude around the asymptotic value
$g(r\rightarrow\infty)=1$. The presence of well defined broad
peaks in the radial distribution function is the hallmark of a
structurally disordered phase, with particles oscillating around
randomly arranged spatial sites. This dynamics is smeared out as
long as the interparticle distance increases, and all correlations
are dramatically reduced after $r\approx 4$, thus indicating the
lack of any long range positional order in the system.
The $g(r)$ reported in Fig. \ref{f8} compares favorably with
previous determination in soft sphere systems.

\section{Conclusions}
We have developed a parallel code which demonstrates the
applicability of molecular dynamics techniques to the study of
billions-body structural glasses. We tested our code on  the
world's largest supercomputer available, namely the Jugene
BlueGene system at the J\"ulich Supecomputing center, and
demonstrated scalability on the full machine [characterized by
$294912$ computing cores] with an efficiency of $89\%$ in the
largest configuration of $100$ billions of particles. We then
applied our code to the case study of the supercooled dynamics of
an exceptionally large system, constituted by an highly frustrated
binary mixture of one billion of particles. This simulation paves the way to the computational study of
billions-body structural glasses, thus achieving new levels of
resolution in the analysis of anomalous vibration of amorphous
materials and, on broader perspective, of living matter.












\end{document}